\documentclass[twocolumn,english,prl,showpacs,floatfix]{revtex4}
\usepackage{amsmath}
\usepackage{amsfonts}
\usepackage{graphicx}
\usepackage{amssymb}
\usepackage{graphicx}


\begin{document}

\title{Atom-atom correlations in colliding Bose-Einstein condensates}
\author{Magnus \"{O}gren}
\affiliation{ARC Centre of Excellence for Quantum-Atom Optics, School of Physical
Sciences, University of Queensland, Brisbane, Queensland 4072, Australia}
\author{K. V. Kheruntsyan}
\affiliation{ARC Centre of Excellence for Quantum-Atom Optics, School of Physical
Sciences, University of Queensland, Brisbane, Queensland 4072, Australia}
\date{\today{}}

\begin{abstract}
We analyze atom-atom correlations in the $s$-wave scattering halo of
two colliding condensates. By developing a simple perturbative
approach, we obtain explicit analytic results for the collinear (CL)
and back-to-back (BB) correlations corresponding to realistic
density profiles of the colliding condensates with interactions. The
results in the short time limit are in agreement with the
first-principles simulations using the positive-$P$ representation
and provide analytic insights into the experimental observations of
Perrin \textit{et al.} [Phys. Rev. Lett. \textbf{99}, 150405
(2007)]. For long collision durations, we predict that the BB
correlation becomes broader than the CL correlation.
\end{abstract}

\maketitle

Experiments with colliding Bose-Einstein condensates (BECs) \cite%
{FWM-exp,Perrin-BEC-collisions} are currently attracting considerable
attention in the field of ultracold quantum gases \cite%
{Julienne,Yurovsky-FWM,Trippenbach2002,Trippenbach,Trippenbach2008,NorrieBallaghGardiner,DeuarDrummond,Moelmer-width,Perrin-theory}%
. A recent breakthrough in this area is the direct detection \cite%
{Perrin-BEC-collisions} of atom-atom pair correlations in the $s$-wave
scattering halo formed in the collision of metastable helium ($^{4}$He$%
^{\ast }$) condensates. Experimental advances like this pose increasingly
demanding challenges to theory due to the need to provide quantitatively
accurate descriptions in realistic parameter regimes.

Theoretical developments are taking place on two fronts using (i)
using numerical techniques based on stochastic phase-space methods
(such as the
first-principles simulations in the positive-$P$ representation \cite%
{DeuarDrummond,Perrin-theory} or approximate simulations based on
the truncated Wigner-function expansion
\cite{NorrieBallaghGardiner}), and (ii) analytic methods such as the Bogoliubov theory \cite%
{Trippenbach2002,Trippenbach,Trippenbach2008}, undepleted source
approximation \cite{Yurovsky-FWM,Perrin-theory}, and a simple
Gaussian ansatz \cite{Moelmer-width}. (The analytic methods often
represent extensions of quantum optics approaches describing closely
related systems of parametric down-conversion and four-wave mixing
\cite{QO}.) Despite these developments, the existing numerical
techniques still fall short of fully
describing the experimental measurements of Ref.~\cite{Perrin-BEC-collisions}%
, whereas the approximations of the analytic methods are usually too severe
to lead to full quantitative agreement with the experiments.
%usually make additional simplifying
%assumptions (e.g., a Gaussian density profile of the colliding condensates)
%which lead to quantitative discrepancies with the experimental observations.

In this paper, we develop an alternative analytic scheme to obtain
explicit results for atom-atom correlations in condensate
collisions. The scheme is rather simple, yet it compares
surprisingly well with the exact positive-$P$ simulations in the
short-time limit. The main advantages of the method are the analytic
transparency and the fact that it can model realistic density
profiles of the colliding condensates with interactions. This is
important for addressing the role of mode mixing due to the
inhomogeneity of trapped atomic clouds, which is the strongest
effect that influences the strengths and the width of atom-atom
correlations. Our results may also have implications for future
experiments aimed at producing relative atom number
squeezing \cite%
{Perrin-theory,Savage,SavageKheruntsyanSpatial,OgrenKheruntsyan} and
Einstein-Podolsky-Rosen correlations \cite{Meystre-spin-EPR,EPR}.

Additionally, we perform first-principles positive-$P$ simulations of the
collision dynamics and analyze the widths of the collinear (CL) and
back-to-back (BB) correlations as per measurements of Ref. \cite%
{Perrin-BEC-collisions}. These characterize, respectively, the pair
correlations between the scattered atoms propagating in the same and
in the opposite directions. The simulations are performed for
$^{23}$Na atoms (instead of $^{4}$He$^{\ast }$ used in
\cite{Perrin-BEC-collisions}) as this case appears to have more
favorable parameters for the positive-$P$ method to remain valid for
long collision durations. For comparison, in the case of
$^{4}$He$^{\ast }$ \cite{Perrin-BEC-collisions,Perrin-theory}, the
simulation durations were much shorter than the experimental
collision time due the smaller mass and larger scattering length of
$^{4}$He$^{\ast }$ atoms. The surprising result that we find here is
that the width of the BB correlation grows with time and eventually
becomes larger than the width of the CL correlation. This
observation is in agreement with the measured correlation widths of
Ref. \cite{Perrin-theory}; it is also accompanied by the reduction
of the BB correlation strength below the CL correlation, which in
turn implies absence of relative number squeezing in the long-time
limit.

We start by considering the equations of motion describing the collision of
two BECs in the Bogoliubov approximation in which the atomic field operator
is split into the mean-field (MF) and fluctuating components, $\widehat{\Psi
}(\mathbf{x},t)=\Psi _{+Q}(\mathbf{x},t)+\Psi _{-Q}(\mathbf{x},t)+\widehat{%
\delta }(\mathbf{x},t)$. Here, $\Psi _{\pm Q}(\mathbf{x},t)$
represent the MF amplitudes of the colliding condensates (which can
be created by splitting a single stationary condensate using a Bragg
pulse or Raman lasers) with mean momenta $+Q$ and $-Q$ along the $x$
axis (in wave number units, in the center-of-mass frame), whereas
$\widehat{\delta }(\mathbf{x},t)$ is the fluctuating component,
which is treated quantum mechanically to the lowest order in
perturbation theory. The fluctuating component satisfies the
equation of motion%
\begin{equation}
i\hbar \frac{\partial \widehat{\delta }(\mathbf{x},t)}{\partial t}\!=\!-%
\frac{\hbar ^{2}\nabla ^{2}}{2m}\widehat{\delta }(\mathbf{x},t)+2U\Psi
\!_{+\!Q}(\mathbf{x},t)\Psi \!_{\!-Q}(\mathbf{x},t)\widehat{\delta }^{\dag }(%
\mathbf{x},t),  \label{x-space-eq-1}
\end{equation}%
where $U=4\pi \hbar ^{2}a/m$ is the coupling constant describing $s$-wave
scattering interactions, with $a$ being the scattering length. The MF
components, on the other hand, satisfy the standard time-dependent
Gross-Pitaevskii (GP) equation. However, if the mean kinetic energy of the
colliding atoms is much larger than the interaction energy and if we
restrict ourselves to short collision times, then the MF amplitudes can be
approximated by freely propagating fields \cite{Trippenbach}: $\Psi _{\pm Q}(%
\mathbf{x},t)=$ $\sqrt{\rho (\mathbf{x})/2}\exp (\pm iQx-i\hbar
Q^{2}t/2m)$. Here, $\rho (\mathbf{x})$ is the density profile of the
initial source condensate and we have additionally assumed that the
center-of-mass displacements $x\mp v_{0}t$ (where $v_{0}=\hbar Q/m$
is the collision velocity) of the MF components are negligible in
the short-time limit. This is justified if the displacements during
the effective collision time are much smaller than the
characteristic size of the colliding BECs in the collision
direction. In the experiments of Ref. \cite{Perrin-BEC-collisions}
this was indeed the case.

Substituting the expressions for $\Psi _{\pm Q}(\mathbf{x},t)$ into Eq.~(\ref%
{x-space-eq-1}) and transforming to a rotating frame $\widehat{\delta }(%
\mathbf{x},t)\rightarrow \widehat{\delta }(\mathbf{x},t)\exp (-i\hbar
Q^{2}t/2m)$, the equation of motion for $\widehat{\delta }(\mathbf{x},t)$
can be rewritten in a simpler form,%
\begin{equation}
\frac{\partial \widehat{\delta }(\mathbf{x},t)}{\partial t}=i\left[ \frac{%
\hbar \nabla ^{2}}{2m}+\frac{\hbar Q^{2}}{2m}\right] \widehat{\delta }(%
\mathbf{x},t)-ig(\mathbf{x})\widehat{\delta }^{\dag }(\mathbf{x},t),
\label{x-space-eq}
\end{equation}%
where $g(\mathbf{x})$ is a spatially dependent effective coupling given by $%
g(\mathbf{x})=2U\Psi _{+Q}(\mathbf{x},0)\Psi _{-Q}(\mathbf{x},0)/\hbar
=U\rho (\mathbf{x})/\hbar $.%
%Equation (\ref{x-space-eq}) is similar to the equation of motion describing
%the evolution of the atomic field in the dissociation of a BEC of molecular
%dimers in the undepleted molecular approximation \cite%
%{Savage,PMFT,OgrenKheruntsyan}.

Converting to Fourier space, $\widehat{\delta }(\mathbf{x},t)=\int d\mathbf{k%
}\widehat{a}(\mathbf{k},t)\exp (i\mathbf{k}\cdot \mathbf{x})/(2\pi )^{3/2}$,
yields the following equation of motion for the amplitude operator $\widehat{%
a}(\mathbf{k},t)$:%
\begin{equation}
\frac{d\widehat{a}(\mathbf{k},t)}{dt}\!=\!-i\Delta _{k}\widehat{a}(\mathbf{k}%
,t)-i\!\int\! \frac{d\mathbf{q}}{(2\pi )^{3/2}}\widetilde{g}(\mathbf{q}+\mathbf{k}%
)\widehat{a}^{\dagger }(\mathbf{q},t).  \label{HeisenbergEquation}
\end{equation}%
Here $\widetilde{g}(\mathbf{k})=\int d\mathbf{x}e^{-i\mathbf{k}\cdot \mathbf{%
x}}g(\mathbf{x})/(2\pi )^{3/2}$ is the Fourier transform of $g(\mathbf{x})$,
$\Delta _{k}\equiv \hbar (k^{2}-Q^{2})/\left( 2m\right) $, and $k^{2}=|%
\mathbf{k}|^{2}$.

From Eq.~(\ref{HeisenbergEquation}) we can easily recognize the role
of mode mixing in the spatially inhomogeneous treatment compared to
an idealized uniform treatment. In the present inhomogeneous case,
the finite
width of the coupling $\widetilde{g}(\mathbf{k})$ implies that $\widehat{a}(%
\mathbf{k})$ couples not only to $\widehat{a}^{\dagger }(-\mathbf{k})$, but
also to a range of momenta around $-\mathbf{k}$, within $-\mathbf{k}\pm
\delta \mathbf{k}$. The spread in $\delta \mathbf{k}$ determines the width
of atom-atom correlations with nearly opposite momenta and is ultimately
related to the momentum width of the colliding BECs.

To quantify the pair correlations expected between the $s$-wave scattered
atoms with equal but opposite momenta due to momentum conservation and
between the atoms in the collinear direction due to quantum statistical
effects we use Glauber's second-order correlation function
\begin{equation}
g^{(2)}(\mathbf{k},\mathbf{k}^{\prime },t)=\frac{\langle \widehat{a}%
^{\dagger }(\mathbf{k},t)\widehat{a}^{\dagger }(\mathbf{k}^{\prime },t)%
\widehat{a}(\mathbf{k}^{\prime },t)\widehat{a}(\mathbf{k},t)\rangle }{n(%
\mathbf{k},t)n(\mathbf{k}^{\prime },t)},  \label{g2-def}
\end{equation}%
which describes the density-density correlations between two momentum
components $\mathbf{k}$ and $\mathbf{k}^{\prime }$. The normalization with
respect to the product of the densities $n(\mathbf{k},t)$ and $n(\mathbf{k}%
^{\prime },t)$ ensures that $g^{(2)}(\mathbf{k},\mathbf{k}^{\prime },t)=1$
for uncorrelated states. The averaging is with respect to the vacuum initial
state for the scattered modes.

The linearity of Eq.~(\ref{HeisenbergEquation}) ensures that one can
apply Wick's theorem to Eq.~(\ref{g2-def}) and factorize the
fourth-order moment, yielding
\begin{equation}
g^{(2)}(\mathbf{k},\mathbf{k}^{\prime },t)=1+\frac{|n(\mathbf{k},\mathbf{k}%
^{\prime },t)|^{2}+|m(\mathbf{k},\mathbf{k}^{\prime },t)|^{2}}{n(\mathbf{k}%
,t)n(\mathbf{k}^{\prime },t)}.  \label{g2-Wick}
\end{equation}%
Here, $n\left( \mathbf{k},\mathbf{k}^{\prime },t\right) =\langle \widehat{a}%
^{\dagger }(\mathbf{k},t)\widehat{a}(\mathbf{k}^{\prime },t)\rangle $ and $%
m\left( \mathbf{k},\mathbf{k}^{\prime },t\right) =\langle \widehat{a}(%
\mathbf{k},t)\widehat{a}(\mathbf{k}^{\prime },t)\rangle $ are the normal and
anomalous densities; $n\left( \mathbf{k},t\right) =n\left( \mathbf{k},%
\mathbf{k},t\right) $ is the momentum distribution.

It follows from Eq.~(\ref{HeisenbergEquation}) and the shape of $\widetilde{g%
}(\mathbf{k})$ that, for sufficiently large collision momentum $Q$
(much
larger than the momentum spread of the colliding BECs), the normal density $%
n\left( \mathbf{k},\mathbf{k}^{\prime },t\right) $ acquires nonzero
population primarily for pairs of nearby momenta, $\mathbf{k}^{\prime
}\simeq \mathbf{k}$, while the anomalous density $m(\mathbf{k},\mathbf{k}%
^{\prime },t)$ does so for pairs of momenta that are nearly opposite, $\mathbf{k}%
^{\prime }\simeq -\mathbf{k}$. Accordingly, we can concentrate on the CL and
BB correlations, which are denoted via $g_{\mathrm{CL}}^{(2)}(\mathbf{k}%
,\Delta \mathbf{k},t)=g^{(2)}(\mathbf{k},\mathbf{k}+\Delta \mathbf{k},t)$
and $g_{\mathrm{BB}}^{(2)}(\mathbf{k},\Delta \mathbf{k},t)=g^{(2)}(\mathbf{k}%
,-\mathbf{k}+\Delta \mathbf{k},t)$ and are given by%
\begin{gather}
g_{\mathrm{CL}}^{(2)}(\mathbf{k},\Delta \mathbf{k},t)=1+\frac{|n(\mathbf{k},%
\mathbf{k+}\Delta \mathbf{k},t)|^{2}}{n(\mathbf{k},t)n(\mathbf{k+}\Delta
\mathbf{k},t)},  \label{CL-corr-def} \\
g_{\mathrm{BB}}^{(2)}(\mathbf{k},\Delta \mathbf{k},t)=1+\frac{|m(\mathbf{k},-%
\mathbf{k+}\Delta \mathbf{k},t)|^{2}}{n(\mathbf{k},t)n(-\mathbf{k+}\Delta
\mathbf{k},t)}.  \label{BB-corr-def}
\end{gather}

To calculate these correlation functions in the short time limit, we proceed
with the Taylor expansion in time, up to the terms of order $t^{2}$ \cite%
{OgrenKheruntsyan}:
\begin{equation*}
\widehat{a}(\mathbf{k},t)=\widehat{a}(\mathbf{k},0)+\left. \frac{\partial
\widehat{a}(\mathbf{k},t)}{\partial t}\right\vert _{t=0}t+\left. \frac{%
\partial ^{2}\widehat{a}(\mathbf{k},t)}{\partial t^{2}}\right\vert _{t=0}%
\frac{t^{2}}{2}+\ldots
\end{equation*}%
The expansion is valid for $t\ll t_{0}$, where $t_{0}=1/g(0)=1/[U\rho
(0)/\hbar ]$ is the time scale \cite{comment1}. Using the right-hand side of Eq.~(\ref%
{HeisenbergEquation}), this gives, up to the lowest-order terms,%
\begin{eqnarray}
n(\mathbf{k},\mathbf{k}^{\prime },t) &\simeq &t^{2}\int d\mathbf{q}%
\widetilde{g}(\mathbf{q}+\mathbf{k})\widetilde{g}(\mathbf{q}+\mathbf{k}%
^{\prime })/(2\pi )^{3}  \notag \\
&=&t^{2}\int d\mathbf{x}e^{-i(\mathbf{k}-\mathbf{k}^{\prime })\cdot \mathbf{x%
}}[g(\mathbf{x})]^{2}/(2\pi )^{3},  \label{normal-moment-b} \\
|m(\mathbf{k},\mathbf{k}^{\prime },t)| &\simeq &t|\widetilde{g}(\mathbf{k}+%
\mathbf{k}^{\prime })|/(2\pi )^{3/2}  \notag \\
&=&t\left\vert \int d\mathbf{x}e^{-i(\mathbf{k}+\mathbf{k}^{\prime })\cdot
\mathbf{x}}g(\mathbf{x})/(2\pi )^{3}\right\vert .  \label{anomalous-moment-b}
\end{eqnarray}%
From these results we see that the width of the CL correlation, Eq.~(\ref%
{CL-corr-def}), is determined by the square of the Fourier transform of the
square of the effective coupling $g(\mathbf{x})$. The width of the BB
correlation, Eq.~(\ref{BB-corr-def}), on the other hand, is determined by
the square of the Fourier transform of $g(\mathbf{x})$. Therefore, the CL
correlation is generally broader than the BB correlation at short times.

\textit{Thomas-Fermi (TF) parabolic density profile.} --- We now give
explicit analytic results for the case when the initial condensate density
profile is given by the ground state of the GP equation in a harmonic trap
in the TF regime: $\rho (\mathbf{x})=\rho
_{0}(1-\sum_{i}x_{i}^{2}/R_{i}^{2}) $ for $\sum_{i}x_{i}^{2}/R_{i}^{2}<1$ ($%
i=x,y,z$) and $\rho (\mathbf{x})=0$ elsewhere, with $R_{i}$ being
the TF radius along direction $i$ and $\rho _{0}$ the peak density.
For definiteness, we consider CL and BB correlations for which the
displacement $\Delta \mathbf{k}$ is along one of the Cartesian
coordinates,
i.e., $\mathbf{k}^{\prime }=\pm \mathbf{k+e}_{i}\Delta k_{i}$, where $\mathbf{%
e}_{i}$ is the unit vector in the $k_{i}$ direction. The integrals in Eqs.~(%
\ref{normal-moment-b}) and (\ref{anomalous-moment-b}) can be performed
explicitly, in terms of Bessel functions $J_{\nu }(z)$ \cite{Bateman},
yielding%
\begin{gather}
n(\mathbf{k},\mathbf{k\!+\!e}_{i}\Delta k_{i},t)\simeq \frac{8t^{2}g^{2}(0)%
\overline{R}^{3}J_{7/2}(\Delta k_{i}R_{i})}{(2\pi )^{3/2}(\Delta
k_{i}R_{i})^{7/2}},  \label{normal-Bessel} \\
|m(\mathbf{k},-\mathbf{k\!\!+\!\!e}_{i}\Delta k_{i},t)|\!\simeq \!\frac{%
2tg(0)\overline{R}^{3}J_{5/2}(\Delta k_{i}R_{i})}{(2\pi )^{3/2}(\Delta
k_{i}R_{i})^{5/2}},  \label{anomalous-Bessel}
\end{gather}%
where $g(0)=U\rho _{0}/\hbar $, and $\overline{R}=(R_{x}R_{y}R_{z})^{1/3}$
is the geometric mean TF radius. Applying these results to $\Delta k_{i}=0$,
using $J_{\nu }(z)\simeq (z/2)^{\nu }/\Gamma (\nu +1)$ for $z\ll 1$ ($\nu
\neq -1,-2,\ldots $), we obtain the following results for the atomic
momentum distribution and the anomalous density: $n(\mathbf{k},t)\simeq
4t^{2}g^{2}(0)\overline{R}^{3}/105\pi ^{2}$ and $|m(\mathbf{k},-\mathbf{k}%
,t)|\simeq tg(0)\overline{R}^{3}/15\pi ^{2}$. These are uniform in the
short-time limit, corresponding to \emph{spontaneous} initiation of the
scattering which populates the scattering modes uniformly \cite%
{Trippenbach,Perrin-theory} without the need to strictly conserve
energy. A narrow scattering shell around $|\mathbf{k}|=Q$ forms
later in time, while the momentum cutoff, which must be assumed for
a $\delta $-function interaction potential, prevents the total atom
number from diverging.

Substituting Eqs. (\ref{normal-Bessel}) and (\ref{anomalous-Bessel}) into
Eqs. (\ref{CL-corr-def})~and~(\ref{BB-corr-def}), and suppressing the
uniform dependency on $\mathbf{k}$, we obtain the following explicit results
for the CL and BB correlations, valid for $t\ll t_{0}$:
\begin{gather}
g_{\mathrm{CL}}^{(2)}(\Delta k_{i},t)\simeq 1+\frac{105^{2}\pi }{2}\left[
\frac{J_{7/2}(\Delta k_{i}R_{i})}{(\Delta k_{i}R_{i})^{7/2}}\right] ^{2},
\label{g2-CL-final} \\
g_{\mathrm{BB}}^{(2)}(\Delta k_{i},t)\simeq 1+\frac{105^{2}\pi }{%
32t^{2}g^{2}(0)}\left[ \frac{J_{5/2}(\Delta k_{i}R_{i})}{(\Delta
k_{i}R_{i})^{5/2}}\right] ^{2}.  \label{g2-BB-final}
\end{gather}

The CL correlation shows the Hanbury Brown and Twiss (HBT) bunching
with the peak value of $g_{\mathrm{CL}}^{(2)}(0,t)=2$. The BB
correlation, on the other hand, shows superbunching
($g_{\mathrm{BB}}^{(2)}(0,t)> 2$) between
atom pairs with equal but opposite momenta, with the peak value $g_{\mathrm{%
BB}}^{(2)}(0,t)=1+7^{2}/[2^{4}t^{2}g^{2}(0)]\gg 1$ partly reflecting
the fact that typical mode occupation numbers in this spontaneous
scattering regime are much smaller than 1. From the above results we
also determine the widths of the CL and BB correlations, which are
defined for simplicity
as the half width at half maximum, $w_{i}^{(\mathrm{CL})}\simeq 1.23w_{i}^{(%
\mathrm{S})}$ and $w_{i}^{(\mathrm{BB})}\simeq 1.08w_{i}^{(\mathrm{S})}$,
giving the ratio of $w_{i}^{(\mathrm{CL})}/w_{i}^{(\mathrm{BB})}\simeq 1.14$%
. Here $w_{i}^{(\mathrm{S})}\simeq 1.99/R_{i}$ is the width of the momentum
distribution $\widetilde{\rho }(k_{i})=|\int d\mathbf{x}\sqrt{\rho (\mathbf{x%
})}\exp (-ik_{i}x_{i})/(2\pi )^{3/2}|^{2}$ of the source condensate along
direction $i$:%
\begin{equation}
\widetilde{\rho }(k_{i})=\frac{\pi \rho _{0}\overline{R}^{6}}{2}\frac{%
\left\vert 2J_{1}(k_{i}R_{i})-k_{i}R_{i}J_{0}(k_{i}R_{i})\right\vert ^{2}}{%
(k_{i}R_{i})^{6}}.
\end{equation}

The ratio of $w_{i}^{(\mathrm{CL})}/w_{i}^{(\mathrm{BB})}\simeq
1.14$ for the TF parabolic density profile (which can be contrasted
with the larger value of $\sigma _{i}^{(\mathrm{CL})}/\sigma
_{i}^{(\mathrm{BB})}=\sqrt{2}$ in the case of a Gaussian profile;
see below) is in good agreement with the results of positive-$P$
simulations of $^{4}$He$^{\ast }$ BEC collisions for
relatively short collision times ($\lesssim 25$ $\mu $s) \cite{Perrin-theory}%
, for which the obtained ratios ranged between $1.08$ and $1.13$
depending on the correlation direction $i$. Similarly, good
agreement is obtained when comparing the individual widths, in which
case the positive-$P$
results were $w_{x}^{(\mathrm{CL})}\simeq 1.27w_{x}^{(\mathrm{S})}$, $%
w_{y,z}^{(\mathrm{CL})}\simeq 1.57w_{y,z}^{(\mathrm{S})}$, $w_{x}^{(\mathrm{%
BB})}\simeq 1.18w_{x}^{(\mathrm{S})}$, and $w_{y,z}^{(\mathrm{BB})}\simeq
1.39w_{y,z}^{(\mathrm{S})}$. The somewhat larger values of the $y,z$%
 correlation widths than the above analytic results are explained by
the
fact that the simulated source condensates were confined in the transverse ($%
y$ and $z$) directions much more strongly than in the axial ($x$)
direction, and therefore the actual GP ground-state profile (which
was used as the initial condition) along $y$ and $z$ was
intermediate between a Gaussian and a TF parabola.

The present analytic results and the positive-$P$ results for $^{4}$He$%
^{\ast }$ \cite{Perrin-theory} are generally in satisfactory agreement with
the experimentally measured absolute widths \cite{Perrin-BEC-collisions},
except that the experimentally measured CL width was somewhat smaller than
the BB width, which is in contrast to the above analytic results and the
positive-$P$ results for $^{4}$He$^{\ast }$. An obvious suspect for this
discrepancy is the fact that the effective collision time in the experiment
was $\sim 150$ $\mu $s, whereas the positive-$P$ simulations of Ref.~\cite%
{Perrin-theory} were restricted to $\lesssim 25$ $\mu $s due to the large
sampling errors developed past that time. Therefore, to explain this
discrepancy it is important to perform first-principle simulations for
longer collision durations and monitor the long-time dynamics of the
correlation widths.

\begin{figure}[tbp]
%\vspace{4.1cm}
\includegraphics[height=3.9cm]{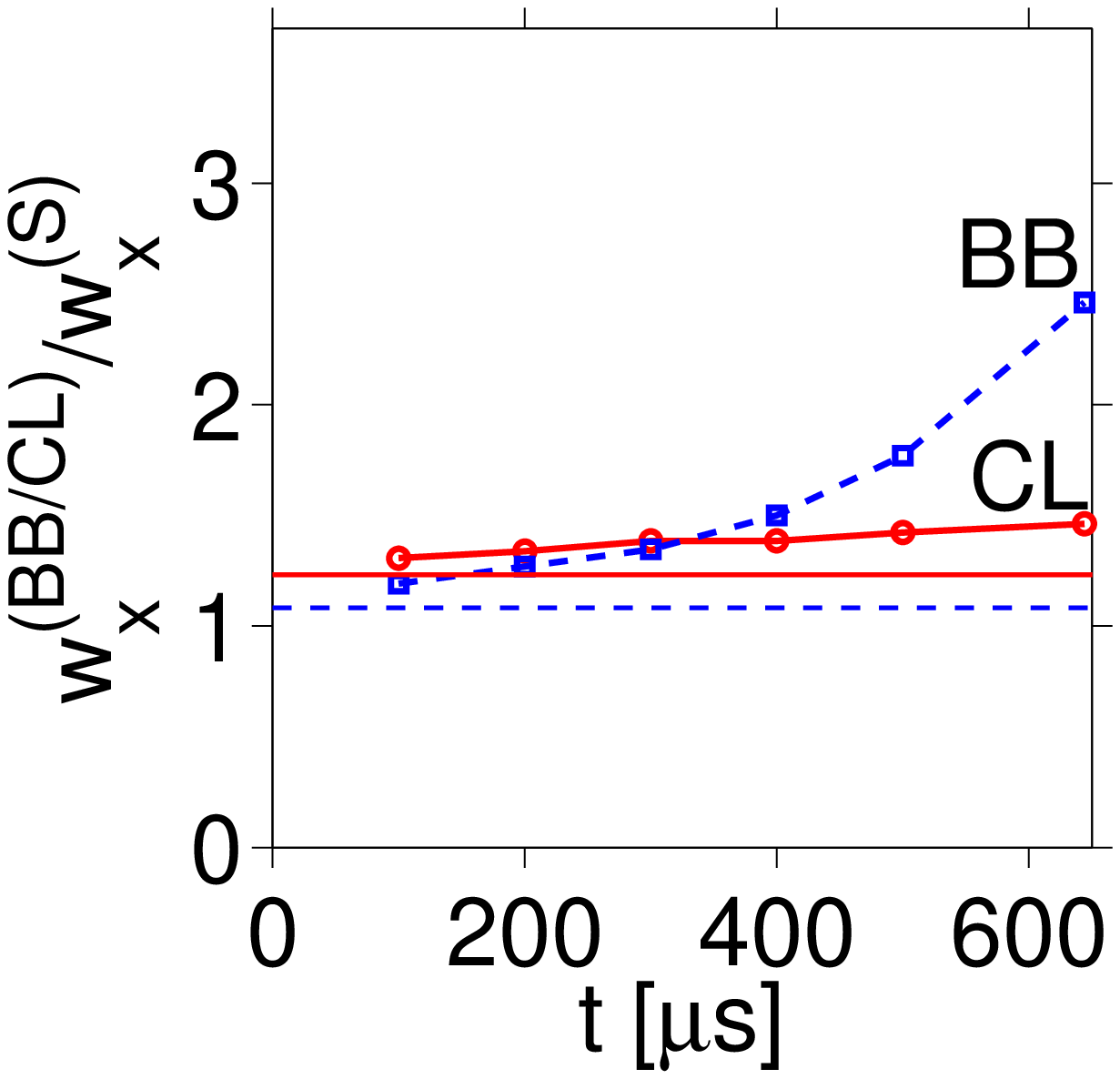}~~~ %
\includegraphics[height=3.9cm]{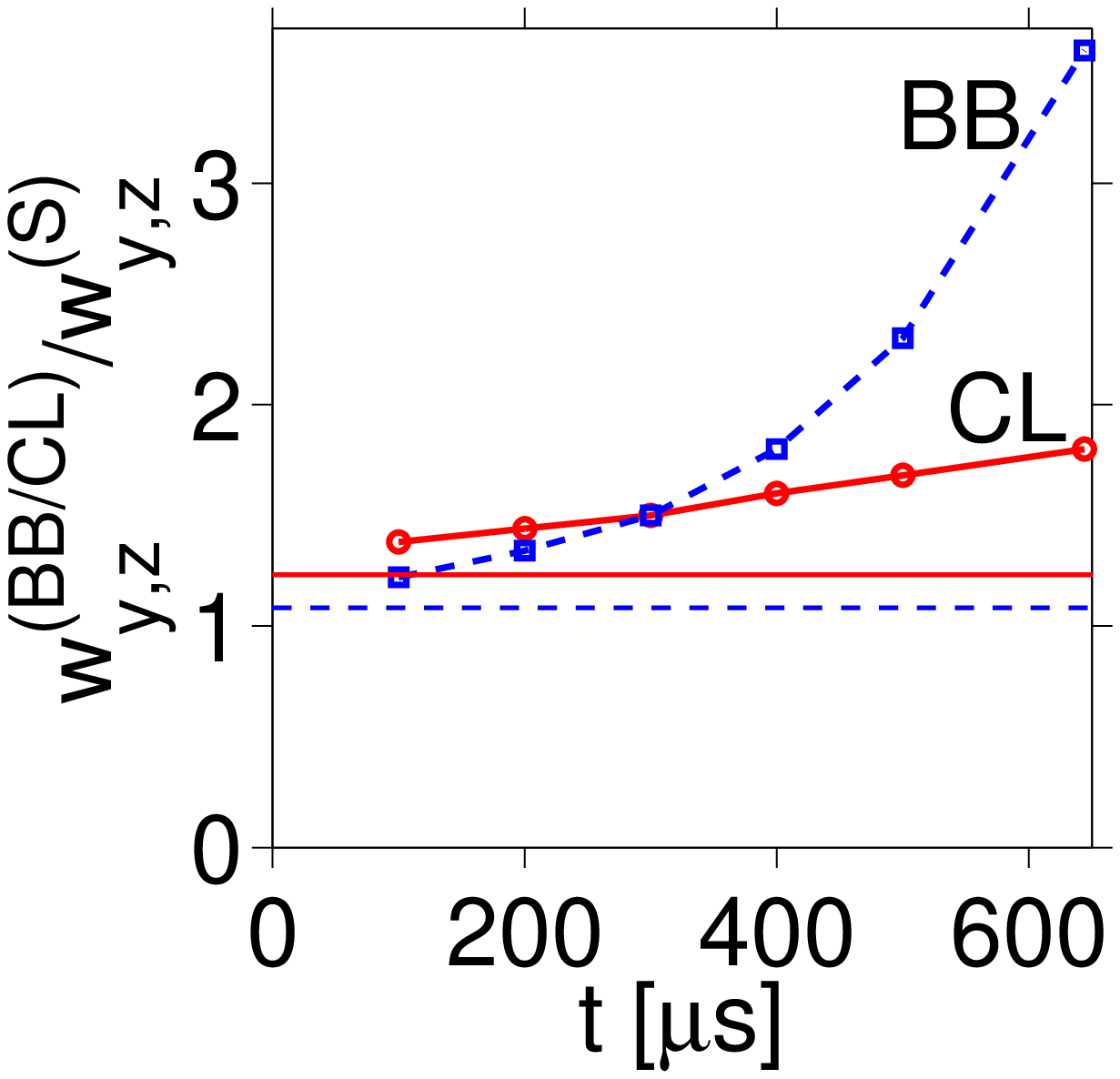}
\caption{Momentum widths of the BB and CL correlations relative to
the width of the source BEC in $x$, $y$, and $z$ directions as a
function of time. The solid and dashed curves with marks are the
results of the positive-$P$ simulations for $^{23}$Na BECs as in
Ref.~\protect\cite{DeuarDrummond}. The horizontal solid and dashed
lines are the corresponding correlation widths from
Eqs.~(\protect\ref{g2-CL-final}) and (\protect\ref{g2-BB-final}).}
\label{figure1}
\end{figure}

To this end we have been able to perform such simulations for collisions of $%
^{23}$Na BECs, in which case the smaller scattering length and larger mass
compared to $^{4}$He$^{\ast }$ give more favorable parameter values for the
positive-$P$ simulations and allow us to extend them to collision durations $%
\lesssim 650$ $\mu $s. More specifically, we have performed the simulations
as in Ref.~\cite{DeuarDrummond}, starting from the full effective
field-theory Hamiltonian, and extracted the CL and BB correlation widths as
a function of time. The results are shown in Fig.~\ref{figure1}, where the
marked solid and dashed curves refer to the numerically obtained CL and BB
widths, respectively. The horizontal solid and dashed lines, on the other
hand, are the corresponding CL and BB widths from the present analytic
treatment, Eqs. (\ref{g2-CL-final}) and (\ref{g2-BB-final}). We see that the
short-time asymptotic limits of the exact numerical results converge to the
analytic predictions, thus proving the validity and usefulness of the
developed analytic approach in this limit.

The long time dynamics of the correlation widths from the positive-$P$
results show that at some point in time the BB correlation width becomes
larger than the CL width. This supports the experimental observations of
Ref.~\cite{Perrin-BEC-collisions} that $w_{y,z}^{(\mathrm{BB}%
)}\!>\!w_{y,z}^{(\mathrm{CL})}$. It is interesting to note that the
crossover point ($\sim \!300$ $\mu $s) is close to the point in time
($\sim \!400$ $\mu $s) where the peak BB correlation drops below the
collinear HBT peak value of $2$ (see Fig.~3(a)
of~\cite{DeuarDrummond}). Both these observations seem to be unique
to the exact first-principle simulations and have not been observed
so far using approximate theoretical techniques. We attribute these
effects to the dynamical broadening of the momentum distributions of
the expanding condensates, and partly to rescattering of scattered
atoms with opposite momenta back into the condensate modes.

The importance of the observation of $g_{\mathrm{BB}%
}^{(2)}(0,t)<g_{\mathrm{CL}}^{(2)}(0,t)$ is related to the prospect
of observing relative number squeezing between the $s$-wave
scattered atoms with equal but opposite momenta
\cite{Perrin-theory}. The relative number squeezing itself is a
manifestation of a violation of the
classical Cauchy-Schwartz inequality, corresponding to $g_{\mathrm{BB}%
}^{(2)}(0,t)>g_{\mathrm{CL}}^{(2)}(0,t)$
\cite{SavageKheruntsyanSpatial}, while the opposite inequality
corresponds to absence of squeezing. While we certainly do have a
violation and relative number squeezing in the short-time
limit \cite{Perrin-theory}, the above observation that $g_{%
\mathrm{BB}}^{(2)}(0,t)$ becomes smaller than $g_{\mathrm{CL}}^{(2)}(0,t)$
implies that the squeezing is lost for long collision durations.

\textit{Gaussian density profile.}~--- We now give the results for a
Gaussian density profile of the source condensate, $\rho
(\mathbf{x})=\rho _{0}\exp (-\sum\nolimits_{i}x_{i}^{2}/2\sigma
_{i}^{2})$, corresponding to the momentum distribution of
$\widetilde{\rho }(k)\propto \exp
(-\sum\nolimits_{i}k_{i}^{2}/2\sigma _{k_{i}}^{2})$, where $\sigma
_{i}$ and $\sigma _{k_{i}}=1/2\sigma _{i}$ are the rms widths.
Following the same
procedures as with the TF parabola, we obtain%
\begin{align}
g_{\mathrm{CL}}^{(2)}(\Delta k_{i},t& \ll t_{0})\simeq 1+\exp (-\Delta
k_{i}^{2}/8\sigma _{k_{i}}^{2}), \\
g_{\mathrm{BB}}^{(2)}(\Delta k_{i},t& \ll t_{0})\simeq 1+\frac{8}{%
t^{2}g^{2}(0)}\exp (-\Delta k_{i}^{2}/4\sigma _{k_{i}}^{2}).
\end{align}%
The correlation widths are now given by $\sigma _{k_{i}}^{(\mathrm{CL}%
)}=2\sigma _{k_{i}}$ and $\sigma _{k_{i}}^{(\mathrm{BB})}=\sqrt{2}\sigma
_{k_{i}}$, resulting in the ratio $\sigma _{k_{i}}^{(\mathrm{CL})}/\sigma
_{k_{i}}^{(\mathrm{BB})}=\sqrt{2}$. These results are in agreement with
those of Ref.~\cite{Moelmer-width} (and Ref.~\cite{Trippenbach} under the
same approximations as here) and represent an alternative derivation in the
Gaussian case.

In summary, we have developed a simple perturbative approach to pair
correlations between the atoms scattered from a collision of two
BECs. The analytic results obtained show how the CL and BB
correlations depend on the shape and the size of the colliding
condensates. The results are compared with exact positive-$P$
simulations and are in good agreement in the short time limit. The
long time dynamics -- accessible through the positive-$P$
simulations of sodium condensates -- give evidence that the BB
correlation width grows with time faster than the CL correlation
width and that it can become broader than the CL width. This
conclusion agrees with the experimental observation of
Ref.~\cite{Perrin-BEC-collisions}, in contrast to previous
theoretical predictions.
%Our analytic approach can be easily generalized
%to describe similar effects in dissociation of a BEC of molecular
%dimers \cite{OgrenKheruntsyan}.

The authors acknowledge support from the Australian Research Council and
thank A. Aspect, C. I. Westbrook, D. Boiron, K. M{\o }lmer, P. Deuar, M.
Trippenbach, P. Zi\'{n}, and J. Chwede\'{n}chuk for stimulating discussions.


\begin{thebibliography}{99}
\bibitem{comment1} With this expansion, the commutation relation is given by
$\left[ \widehat{a}(\mathbf{k},t),\widehat{a}^{\dagger }(\mathbf{k}^{\prime
},t)\right] \approx \delta (\mathbf{k}-\mathbf{k}^{\prime })$ up to terms of
order $t^{2}$.
\end{thebibliography}
\end{document}